\documentstyle[aaspp4,epsf]{article} 

\begin{document}
\title{Vegetation's Red Edge: A Possible Spectroscopic Biosignature of Extraterrestrial Plants} 
\author{S. Seager}
\affil{Department of Terrestrial Magnetism, Carnegie Institution of Washington,
5241 Broad Branch Rd. NW, Washington, DC 20015}
\author{E.L. Turner, J. Schafer, E.B. Ford\footnote{Present address: Astronomy Department, 601 Campbell Hall, University of California at Berkeley,
Berkeley, CA 94720-3411}}
\affil{Princeton University Observatory, Princeton, NJ 08540}

\begin{abstract}
Earth's deciduous plants have a sharp order-of-magnitude increase in
leaf reflectance between approximately 700 and 750 nm wavelength.
This strong reflectance of Earth's vegetation suggests that surface
biosignatures with sharp spectral features might be detectable in the
spectrum of scattered light from a spatially unresolved extrasolar
terrestrial planet. We assess the potential of Earth's
step-function-like spectroscopic feature, referred to as the ``red
edge", as a tool for astrobiology.  We review the basic
characteristics and physical origin of the red edge and summarize its
use in astronomy: early spectroscopic efforts to search for vegetation
on Mars and recent reports of detection of the red edge in the
spectrum of Earthshine (i.e., the spatially integrated scattered light
spectrum of Earth).  We present Earthshine observations from Apache
Point Observatory to emphasize that time variability is key to
detecting weak surface biosignatures such as the vegetation red edge. We
briefly discuss the evolutionary advantages of vegetation's red edge
reflectance, and speculate that while extraterrestrial ``light
harvesting organisms'' have no compelling reason to display the exact
same red edge feature as terrestrial vegetation, they might have
similar spectroscopic features at different wavelengths than
terrestrial vegetation.  This implies that future
terrestrial-planet-characterizing space missions should obtain data
that allow time-varying, sharp spectral features at unknown
wavelengths to be identified. We caution that some mineral
reflectance edges are similar in slope and strength to vegetation's
red edge (albeit at different wavelengths); if an extrasolar planet
reflectance edge is detected care must be taken with its interpretation.

\end{abstract}

\section{Introduction}

The search for extrasolar terrestrial planets is in large part
motivated by the hope of finding signs of life or habitability via
spectroscopic biosignatures.  Spectroscopic biosignatures are spectral
features that are either indicative of a planetary environment that is
hospitable to life (such as the presence of liquid water) or of strong
indicators of life itself (such as abundant O$_2$ in the presence of
CH$_4$). Most attention so far has been given to atmospheric
biosignatures, gases such as O$_2$, O$_3$, H$_2$O, and CH$_4$ (Des
Marais et al. 2002).  Instead we focus on a potential {\it surface
biosignature}.  An ideal surface biosignature would be produced by a large
and abrupt change in the reflectance at wavelengths that penetrate to
the planetary surface. Earth has one such surface biosignature: the
vegetation red edge spectroscopic feature
(Figure~\ref{fig:plantspectrum}).

Over one hundred extrasolar giant planets are currently known to orbit
nearby sun-like
stars\footnote{http://exoplanets.org/}$^,$\footnote{http://obswww.unige.ch/$\sim$udry/planet/coralie.html}$^,$\footnote{http://cfa-www.harvard.edu/planets/}(e.g.,
Butler et al. 2003; Mayor et al. 2003).  These planets have been
detected by the radial velocity method, which measures the star's
line-of-sight motion due to its orbit about the planet-star common
center of mass. Therefore, with the exception a half-dozen known
transiting planets, only the minimum mass and orbital parameters are
known.  Many efforts are underway to learn more about extrasolar
planets' physical properties from ground-based and space-based
observations and via proposed or planned space missions.  Direct
detection of scattered or thermally emitted light from the planet
itself is the only way to learn about many of the planet's physical
characteristics. Direct detection of Earth-size planets, however, is
extremely difficult because of the extreme proximity in the sky of the
parent star that is typically $10^6$ to $10^{10}$ times brighter than
the planet.

Terrestrial Planet
Finder\footnote{http://planetquest.jpl.nasa.gov/TPF/tpf\_index.html}
(TPF) is being planned by NASA to find and characterize terrestrial
planets in the habitable zones\footnote{The habitable zone is defined
as the annulus around the star where the planet equilibrium
temperature range is consistent with surface liquid water (Kasting,
Whitmire, \& Reynolds 1993).} of nearby stars.  NASA is currently
planning both a visible-wavelength mission (launch date 2014) and a
mid-infrared-wavelength mission (launch date 2019) for TPF. The ESA
Darwin\footnote{http://sci.esa.int/home/darwin/index.cfm} mission at
mid-infrared wavelengths has similar science goals. The primary focus
for both of these space missions is the direct detection of extrasolar
terrestrial planets and the spectroscopic characterization of their
atmospheres. In addition to the previously mentioned atmospheric
biosignatures, O$_2$, O$_3$, H$_2$O, and CH$_4$, other atmospheric
features such as CO$_2$ and Rayleigh scattering as well as physical
characteristics such as temperature and planetary radius could be
constrained from low-resolution spectra (see e.g., Seager 2003).
Beyond TPF, NASA is contemplating a series of increasingly ambitious
astrobiology missions with evocative names such as Life Finder and
Planet Imager.

New ideas are being developed to maximize science return from TPF.  We
have shown (Ford, Seager \& Turner 2001) that several important
planetary characteristics could also be derived from photometric
measurements of the planet's time domain variability at visible and
near infrared wavelengths. A time series of photometric data of a
spatially unresolved Earth-like planet could reveal a wealth of
information such as the existence of weather, the planet's rotation
rate, presence of large oceans or surface ice, and existence of
seasons.  This would be possible for other planets if the observed
light is scattered by at least two different scatterers which have
significant differences in their albedo, color, or directionality of
scattering.  The amplitude variation of the time series depends on
cloud-cover fraction; more cloud cover raises the average albedo of
the planet and makes it more photometrically uniform, thus
reducing the fractional variability. The signal-to-noise necessary for
photometric study would be obtained by any mission capable of
measuring the sought-after atmospheric biosignature spectral
features. Furthermore, the photometric variability could be monitored
concurrently with a spectroscopic investigation, as was done for the
transiting extrasolar giant planet HD209458b (Charbonneau et
al. 2002).

Serious attention to the possibility of spectroscopic detection of
surface biosignatures on extrasolar plants is relatively recent, even
though interest in spectroscopic signatures of vegetation on planets
in the Solar System has a long history (see Section 3).  The
pioneering detection of the Earth's vegetation red edge feature in the
spectrum of Earthshine by Woolf et al. (2002) and Arnold et
al. (2002) has given the topic of surface biosignatures an empirical
basis (see Section 3 for more details).  In addition to the Woolf et
al. (2002) and Arnold et al. (2002) observational studies, there have
been theoretical discussion papers about surface
biosignatures. Wolstencroft \& Raven (2002) give a detailed discussion of
the likelihood that O$_2$-producing photosynthesis will develop on
Earth-like extrasolar planets.  They consider chemical, astrophysical,
climatological, and evolutionary processes to explore extrasolar
photosynthetic mechanisms and detectability. They conclude that,
despite many uncertanties and alternate possibilities, photosynthesis
is likely to develop on many Earth-like planets and that its mechanism
might well be identical to terrestrial photosynthesis. The exception
is planets orbiting stars much cooler than the Sun; in this case they
postulate mechanisms able to employ lower energy photons via multiple
photon photochemical processes would still allow photosynthesis, 
with the 3- or 4-photon process photosynthesis implying a red-edge at
longer wavelengths.  Due primarily to skepticism about the
practicality of a TPF-type mission in the optical and near-IR spectral
bands, they conclude that detection of a red edge spectroscopic
feature is impractical and that atmospheric oxygen is a much more
promising photosynthesis biosignature.  Knacke (2003) focuses on the
spectroscopic detection of microbial life.  Single-celled microbial
life is the dominant portion of Earth's biomass
today\footnote{Microbial cell numbers on Earth are estimated to be on
the order of $5\times10^{30}$, with the total amount of carbon in
these cells equal to that in all plants on Earth, and the total
nitrogen and phosphorus content 10 times that of all plant biomass
(Madigan et al. 2002).}  (Madigan, Martinko, \& Parker 2002) and for
the majority of time of life on Earth microbial life was the only life
on Earth.  On Earth microbe colonies form large aggregates in oceans
(and sometimes on land); algae and plankton colonies have
spectroscopic features due to chlorophylls and related photosynthetic
pigments.  Knacke (2003) points out, however, that visible wavelength
microbial signatures are confused by the much stronger atmospheric
reflectance, and any red edge feature redward of 600 to 700 nm would
be extremely weak due to the high opacity of sea water at those
wavelengths. Techniques to detect microbial life will be essential if
microbial life is common but multi-cellular life is rare in
extraterrestrial environments.  The best chance for detection and
study of microbes will be with post-TPF/Darwin type missions.

For the purposes of this paper we assume that it will be possible to
carry out a TPF-like space mission at visible to near IR wavelengths
either in the next decade or at some later time. We believe
that searching for spectroscopic signatures of light harvesting
organisms, ({\it i.e.}, spectroscopically resembling terrestrial
plants) is an exciting science goal that will be carried out for at
least some TPF targets.  

The vegetation red edge feature's relevance and usefulness as a
surface biosignature for astrobiological studies remain open to question
and are the topics of this paper (see also the conference
proceedings by Seager \& Ford (2004) for a preliminary discussion.)  In
Section 2 of this paper we summarize relevant current knowledge of the
red edge as it occurs in terrestrial vegetation, including its
physical origin. Section 3 briefly reviews early attempts to search for
vegetation on Mars via spectroscopic signatures and recent reports of
detection of the red edge in Earthshine.  The role of temporal
variability in the detection of surface biosignatures is described in
Section 4.  Observations of the Earthshine feature are also reported
in Section 4, including temporal behavior that supports the
identification of the observed spectroscopic feature with plants on
the Earth's surface.  In Section 5 we speculatively, but critically,
consider the likelihood that the same red edge feature seen on Earth
would occur for vegetation on another world.  In Section 6 we
point out the possibility of confusion of the vegetation red edge
reflectance feature with mineral edge reflectance features. Finally,
we draw some tentative conclusions in Section 7.

\section{Earth's Vegetation Red Edge Spectral Feature}

\subsection{Empirical Characteristics}

Land-based chlorophyll-producing vegetation has a very strong rise in
reflectivity at around 700~nm by a factor of five or more.  This
``red-edge'' spectral signature is much larger than the familiar
chlorophyll reflectivity bump at 500~nm, that gives vegetation its
green color.  Figure~\ref{fig:plantspectrum} shows a deciduous plant
leaf reflection spectrum (Clark et al. 1993). The high absorptance at
UV wavelengths (not shown) and at visible wavelengths is by
chlorophyll and is used by the leaf for photosynthesis. Photosynthesis
is the process by which plants and some other organisms use energy
from the sun to convert H$_2$O and CO$_2$ into sugars and O$_2$.  The
primary molecules that absorb the light energy and convert it into a
form that can drive this reaction are chlorophyll A (absorption maxima
in diethyl ether at $\sim 430$~nm and $\sim 662$~nm) and
chlorophyll B (absorption maxima in diethyl ether at $\sim 453$~nm
and $\sim 642$~nm).

A deciduous plant leaf is strongly reflective between 700~nm and
1000~nm.\footnote{If human vision were sensitive a little further
toward the red, the natural world would be a very different in
appearance: plants would be very red and exceedingly bright (see
Figure~\ref{fig:plantcurve}).}  Figure~\ref{fig:plantcurve} shows that
the leaf also has a very high transmittance at these same wavelengths,
such that reflectivity plus transparency is near 100\% at near-IR
wavelengths. Remarkably, the bulk of the energy in solar radiation
(Figure~\ref{fig:atm}) at sea level is at approximately 600 to
1100~nm---the same wavelength region where the deciduous leaf
reflects or transmits almost all of the solar radiation (Gates et
al. 1965).  The exact wavelength and strength of the red edge depends
on the plant species and environment.  Although negligible from the
biosignature detection view point, it is interesting to note that the
specific wavelength dependence and strength of the red edge feature is
used for remote sensing of specific locations on Earth to identify
plant species and also to monitor a field of vegetation's (such as
crops) health and growth as the red edge changes during the growing
season. It is also useful to note that even conifers and desert plants
have similar red edge features (see Clark et al. 1993).

In the near-infrared (Figure~\ref{fig:plantspectrum}),
plants also have water absorption bands. The band strength depends on
plant water content, weather conditions, plant type, and geographical
region.  These absorption features can be strong, on the order of 20\%
for the water bands at 1.4 and 1.9 $\mu$m, but are not very useful for
identifying life, since they would only be indicative of water and
would likely not be distinguishable from atmospheric water vapor.

Ocean plankton blooms near coastal shores are colorful and large-scale
(e.g., Behrenfeld et al. 2001; Knacke 2003).  The color change in
plankton blooms are due to the increased presence of
chlorophyll---causing the presence of the 500~nm reflectance
peak. Unfortunately, the change in intensity is small for a spatially
unresolved extrasolar planet, especially when considering the low
reflectivity of Earth's oceans and the small area of plankton blooms
compared to a large part of a hemisphere.

\subsection{Physical Nature} 

A typical plant leaf spectrum exhibits the two different
behaviors of leaves: a strong absorption in the visible, and high
scattering (transmission + reflection) in the very near IR (750-1300
nm), sometimes called the ``red edge'' or ``infrared plateau'' by
vegetation remote sensing experts (Figure~\ref{fig:plantcurve}).
The large-scale physical structure of leaves (the layers of cells
and gaps between them) promotes scattering and therefore within a
plant leaf absorption only occurs at wavelengths where specific
chemical pigments or molecules absorb light.  The visible-wavelength
absorbers are determined to be chlorophylls or other pigments
experimentally: in ``variegated'' plant leaves (leaves that are both
green and white) the white part of the leaf has a reflectance as high
in the visible as at near-IR wavelengths; and chlorophyll pigments are
very absorptive at visible wavelengths but are not at all absorptive
at near-IR wavelengths (Knipling 1970 and references therein). Water
absorbs portions of the near-IR reflectance spectrum.  The wavelengths
of water absorption bands are well known and confirmation of water as
the near-IR absorber has been shown with dehydrated leaves which have
a much higher near-IR reflectance spectra than hydrated leaves
(Knipling 1970).

Plant leaves are highly reflective at near-IR wavelengths; few
substances in nature reach this level of reflectivity.  The reason a
plant leaf is so highly reflective (and transmittive) is due to the
leaf construction---structures at all scales make light scattering
highly efficient (see e.g., Gates et al. 1965; Knipling 1970).  
The inside of a leaf is made up of water-filled cells with air gaps
surrounding the cells.  Light reflects off of the cell walls but also
refracts through cell walls from the surrounding air gaps between
cells. Inside the cell, light keeps scattering (except at
pigment-absorbing wavelengths) until it exits the cell. Two causes
keep light scattering inside the cell.  Primarily, the high change in
the index of refraction (from 1.33 for water to 1.00 for air) makes
efficient internal reflection inside the water-filled cells at the
interface between cell walls and the surrounding air
gaps. Secondarily, light will Mie or Rayleigh scatter off of the
cell's organelles which have sizes on the order of the wavelength of
light. Conclusive evidence of the importance of the internal
reflection to the plant leaf's high reflectance was shown by Knipling
(1970). A water infiltrated leaf with water filling the cavities
between cells to form a continuous liquid water medium in the plant
leaf, has a high transmittance at the expense of lower reflectance. In
other words, without internal reflection inside cells, most of the
light entering the top of the leaf would travel down and exit out of
the bottom leaf surface, but when cells are surrounded by an air gap a
significant fraction of the incoming light is redirected upwards by
repeated internal reflections.  The leaf surface contributes only a
small amount to reflectance (Knipling 1970) by either specular
reflection or scattered light. The amount of reflection depends on the
presence of leaf surface wax (increasing reflection at visible (e.g.,
Grant 1987) and near-IR wavelengths (e.g., Slaton et al. 2001)) and
distribution, size, shape and angles of any hairs present (either
increasing or decreasing reflectance; Grant 1987).  Only at large
angles of incidence is radiation predominantly reflected by leaf
surface specular reflection compared to leaf interior reflection
(Grant 1987).

Eventually, the radiation will scatter out of the leaf; designated
``reflected'' at the top or ``transmitted'' at the bottom.  To support
the explanation that radiation is scattered or transmitted everywhere
except at wavelengths with absorption, Figure~\ref{fig:plantcurve}
shows that a leaf's reflectance and transmittance spectrum are
very similar.  The magnitude, wavelength dependence, and ratio of the
reflectance and transmittance is a complex function of the cell size
and shape and the size and shape of the air gaps between the cells
and hence depends on plant type.
3D Monte Carlo simulations are able to reproduce reflection and
transmission spectral properties of plant leaves (e.g., Govaerts et
al.  1996).

The overall reflectance of a plant is lowered by the plant canopy.
The canopy effects include leaf orientation, shadows, and non foliage
background surfaces such as soil (Knipling 1970).  Leaf orientation,
or specifically the angle of incidence and reflectance of radiation,
is important since leaves scatter light anisotropically.  Reflection
at visible wavelengths is reduced by the canopy by approximately twice
as much ($\sim$40\% of a single leaf) as reflection at near-IR
wavelengths ($\sim$70\% of a single leaf) (Knipling 1970). This effect
actually enhances the red-edge amplitude and is due to the near-IR
incident radiation reflecting off of lower leaves and retransmitting
upward through upper leaves.

The high ratio of absorbed to scattered (reflected + transmitted)
radiation on either side of the red edge is not fully understood. One
explanation uses environmental adaptation. If plants absorbed solar
radiation with the same efficiency at longer wavelengths than the red
edge compared to visible wavelengths, then the plants would become too
warm and the proteins irreversibly damaged (Gates et al. 1965). Gates
\& Benedict (1963) have shown that approximately 75\% of the total
energy absorbed by plants is reradiated, while approximately 25\% is
dissipated by convection and transpiration. Therefore, although
thermal regulation is partially controlled by leaf stomata and water
vapor content, radiation balance must also play a role in thermal
regulation. It is likely that a plant balances the competing
requirements of absorption of sunlight at wavelengths appropriate for
photosynthesis reactions with efficient reflectance at other
wavelengths to avoid overheating (Gates et al. 1965). Alternate to the
environment adapation argument is an argument that the cell spacings
that cause the high near-IR reflectance evolved for reasons other
than thermal balance.  The large intercelluar spaces aid gas exchange
(Konrad et al. 2000) and increase the absorption of photosynthetically
active radiation (DeLucia 1996). The thermal regulation argument is
further weakened by the belief that higher plants evolved from aquatic
ancestors; aquatic vegetation is less subject to overheating due
to close thermal coupling with the ambient water.

\section{Red Edge Observations within the Solar System}
\subsection{Spectroscopic Searches for Vegetation on Mars}

During the early part of the last century, speculation about
vegetation on Mars was fueled by reports of a wave of darkening that
appeared during many martian springs (Lowell 1904; Sinton 1958 and
references therein). This wave of darkening changed large dull colored
regions to darkish green hues and proceeded from pole to equator
within a few weeks of the disappearance of the polar caps. The
proposed explanation was that the vegetation was nourished by the
melting polar snow. Although other explanations were put forth, for
proponents of life on Mars the vegetation theory was second in
popularity only to the martian canals. Because the spectrum of
vegetation (Figure~\ref{fig:plantspectrum}) is different from other
green materials, remote sensing was used to test the Earth-like
vegetation hypothesis. By taking spectra of different areas of Mars
with a large telescope (72''), Millman (1939) was able to rule out
Earth-like vegetation based on measurements of the absence of a 500~nm
chlorophyll bump. Others used the red edge reflectance to rule out the
vegetation hypothesis (Slipher 1924; Tickhov 1947; Kuiper 1952). Later
it was shown that the darkening on Mars was caused by changes due to
windblown dust (Sagan and Pollack 1969 and references therein).

In the 1950s new excitement and controversy arose from Sinton's (1957)
claim of evidence for vegetation on Mars. Sinton observed near-IR
absorption features of organic molecules that were also observed to be
present in lichen and some dried plants.  Specifically, the C-H
vibration bands occur in the 3 to 4 $\mu$m region.  Sinton argued that
Mars may have vegetation similar to the hardiest vegetation on Earth:
lichens (a symbiosis of fungi and algae). Lichens show the C-H near-IR
absorption bands but lack the red-edge vegetation spectrum even though
they carry out photosynthesis.  This theory of vegetation on Mars was
ruled out in the early 1960s when two of the observed C-H absorption
bands were identified as deuterated water in Earth's atmosphere (Rea
et al. 1965).

\subsection{Observing Earth as a Reference Case Extrasolar Planet}

Because Earth is the only planet known to harbor life it is the
obvious and only test case of techniques for the search for life on
extrasolar planets.

Sagan et al. (1993) used the Galileo spacecraft for a ``control
experiment'' to search for life on Earth using only conclusions
derived from data and first principle assumptions. En route to
Jupiter, the Galileo spacecraft used two gravitational assists at
Earth (and one at Venus). During the December 1990 fly-by of Earth,
the Galileo spacecraft took low-resolution spectra of different areas
of Earth. In addition to finding ``abundant gaseous oxygen and
atmospheric methane in extreme thermodynamic disequilibrium'', Sagan
et al. (1993) found ``a widely distributed surface pigment with a
sharp absorption edge in the red part of the visible spectrum'' that
``is inconsistent with all likely rock and soil types''. Observing
$\sim$100 km$^2$ areas of Earth's surface the vegetation red edge
feature showed up as a reflectance increase of a factor of 2.5 between
a band centered at 670~nm and one at 760~nm. In contrast
there was no red-edge signature from non-vegetated areas.

A new technique for observing the Earth as an extrasolar planet test
case is now emerging: using Earthshine to study the spatially
unresolved Earth. These studies are similar to Sagan et al.'s
(1993), but they utilize Earth's spatially integrated light as if
Earth were a point source, instead of areas of 100~km$^2$.  Earthshine
is sunlight that has been scattered by the Earth toward the Moon and
then back to Earth. It is often visible as a faint glow on the
otherwise dark region of the lunar disk during crescent
phases. Earthshine can be studied with a CCD camera and specialized
coronagraph even as the Moon waxes (Goode et al. 2001). The very rough
lunar surface makes the Moon a diffuse reflector with each point on
the Moon reflecting the spatially integrated illumination from the
Earth.  The viewing geometry of Earth is the sunlit illuminated
portion of the Earth as seen from the Moon.  A prescient paper by
Arcichovsky (1912) proposed looking for vegetation signatures in
Earthshine as a reference case for vegetation or chlorophyll searches
on other planets. So, while activity and progress in Earthshine
research is recent; the idea itself is an old one.

Earthshine observations can provide a valuable complement to
existing satellite data sets that have been obtained for a wide
variety of purposes and with diverse instruments.  Since most
satellite data are collected by looking nearly straight down at
limited regions of the Earth's surface and most often with relatively
direct solar illumination angles, derivation of the integrated
spectrum of the unresolved Earth with its varying atmospheric path
lengths and solar illumination angles is not generally possible.  Moreover,
many satellite observations are obtained without attention to absolute
flux calibration (because it is not necessary to the purposes for
which the data were obtained) and thus cannot be combined with other
data to derive an integrated spectrum.  Thus, for the purpose of time
domain comparisons of the unresolved Earth's spatially integrated spectrum,
Earthshine provides the best currently available information.


Recent spectral observations of Earthshine have tentatively detected
the red edge feature at the 4 to 10\% level.  Woolf et al. (2002)
observed the setting crescent moon from Arizona which corresponds to
Earth as viewed over the Pacific Ocean. Nevertheless their spectrum
(Figure~2 in Woolf et al. 2002) shows a tantalizing rise just redward
of 700~nm which they tentatively identify with the red edge
feature. The same figure also shows other interesting features of
Earth's visible-wavelength spectrum, notably O$_2$ and H$_2$O
absorption bands (note that spectral lines of both O$_2$ and H$_2$O
cut into the red-edge signature.)  Arnold et al. (2002) made
observations of Earthshine on several different dates. With
observations from France the Earthshine on the evening crescent moon
is from America and the Atlantic Ocean whereas the Earthshine on the
morning crescent moon is from Europe and Asia. After subtracting
Earth's atmospheric spectrum to remove the contaminating atmospheric
absorption bands, they find a vegetation red edge signal of 4 to 10\%.
However, note that it will be extremely difficult to subtract an
unknown extrasolar planet's atmospheric spectrum in the kind of
low-signal-to-noise data expected with future space missions TPF
and Darwin.

An alien civilization observing the spatially unresolved Earth 
with a TPF-like telescope would have difficulty in identifying the
vegetation red-edge signature in medium spectral resolution
data. While the red-edge spectroscopic feature is very strong for an
individual plant leaf, at a factor of five or more, it is much
reduced---down to several percent---when averaged over a (spatially
unresolved) hemisphere of Earth.  This is due to several effects,
including extremely large vegetation-free regions of the Earth's
surface, the presence of clouds that block sight lines to the
surface, anisotropic scattering by vegetation canopies, and soil
characteristics. In addition, the reflectance of vegetation is
anisotropic, so the illumination conditions and planet viewing angle
are important. Nevertheless, at a signal level of a few to several
percent Earth's vegetation red edge is still a viable surface
biosignature for a TPF-like mission.  In favorable
hypothetical cases, however, the feature could be much stronger than
the ones so far observed in Earthshine.  For example, if the planet
cloud cover fraction is lower, a larger fraction of the surface is
covered by vegetation, or a smaller fraction of the illuminated
portion of the planet is visible ({\it i.e.}, a crescent phase), the
fractional amplitude of the red edge would be increased in the
integrated spectrum.

In contrast to the red edge signature, the chlorophyll bump at
500~nm is extremely small in a hemispherically averaged spectrum
of the Earth.  The spectral signature of oceanic vegetation or
plankton is also unlikely to be detectable with TPF/Darwin generation
telescopes, due to strong absorption by particles in the water and
the strong absorptive nature of liquid water at red
wavelengths (see Knacke 2003). 

In the next section we argue that temporal variability may be key for
detection of weak but changing surface features such as the vegetation
red edge feature.

\section{Using Temporal Variability to  Detect Surface Biosignatures}

A time series of data in different bands should help make it possible
to detect a small but unusual spectral feature, even with variable
atmospheric features.  As the continents rotate in and out of view, a
planet's brightness, colors and spectrum will change.  Most of Earth's
surface features, such as sand or ice, have a continuous or minimal
change in albedo with wavelength, in contrast to the abrupt
vegetation red edge spectral feature.

We have shown (Ford, Seager, \& Turner 2001) via simulations that the
existence of different surface features on a planet may be discernible
at visible wavelengths as different surface features rotate in and out
of view. Considering a cloud-free Earth, the diurnal flux variation
caused by different surface features could be as high as 200\%.  This
high flux variation is not only due to the high contrast in different
surface components' albedos, but also to the fact that a relatively
small part of the visible hemisphere dominates the total flux from a
spatially unresolved planet.  Clouds increase the Earth's albedo,
interfere with surface visibility, and reduce the fractional amplitude
of the diurnal light curve to roughly $\sim$10--20\%.  Because cloud
cover is stable for days over some regions, the rotational period of
Earth could still be measured (Ford et al. 2001).

We here restrict our attention to diurnal time scales. However, any
vegetation indicator might also display seasonal (on an orbital
timescale) variations. In this section we describe Earthshine
observations from two different Earth--Moon geometries made to detect
temporal variation of the vegetation red edge signature.

\subsection{APO Earthshine Observations}
\label{sec-observations}
Spectra of Earthshine were obtained with the Double Imaging
Spectrograph (DIS) on the Apache Point Observatory 3.5-meter
telescope\footnote{Based on observations obtained with the Apache
Point Observatory 3.5-meter telescope, which is owned and operated by
the Astrophysical Research Consortium.}  between 11:30 and 13:00 on 8
Feb 2002 (UT) with the Moon rising in the eastern sky and again
between 01:00 and 01:30 on 16 Feb 2002 (UT) with the Moon setting in
the western sky.  As shown in Figure~\ref{fig:apo}, the dark side of
the Moon was illuminated by light scattered from quite different
portions of the Earth's surface on the two dates.  About 20\% of the
visible portion of the Moon's surface was illuminated by sunlight on
the 8 Feb 2002 and about 10\% on 16 Feb 2002.  Airmass was necessarily
high for both sets of observations, ranging from approximately 3 to 7,
the latter being the approximate elevation limit of the telescope.
The same spectroscopic setup was used on both dates; in particular a
150 lines/mm red blazed grating was used to obtain spectra of
approximately 10~${\rm \AA}$ resolution with a 1.2 arcsec by 5 arcmin ``bar
slit''.  Although the whole range from about 400 to 1000~nm was
observed, we here concern ourselves only with the 600-900~nm
range, roughly centered on the red edge feature and in which a
relatively clean and high signal-to-noise spectrum could be extracted.
The detector was an unthinned, and thus relatively unaffected by CCD
fringing, 800x800 TI device.

During the 8 Feb 2002 observations, 8 exposures were obtained of the
dark limb of the Moon ({\it i.e.}, the Earthshine illuminated limb),
followed by 3 exposures of a sunlit mountain peak just on the dark
side of the lunar terminator and then followed by 7 more exposures on
the dark limb.  On 16 Feb 2002, due to a much shorter total available
observing time, only 4 dark limb and 1 sunlit mountain top exposures
were obtained.  Dark limb exposures were made with the slit
approximately centered on and perpendicular to the limb, so as to
obtain both an Earthshine signal and a sky spectrum.  Typical exposure
times were 60 sec on the dark limb and 10 sec on the sunlit mountain
peaks.  Exposures with the ``slit viewer" target acquisition camera
indicated that it would be impossible to obtain DIS exposures of the
fully sunlight illuminated bright limb without saturating the
detector, even with the shortest available shutter times; the
availability of small (spatially unresolved) areas of sunlit lunar
surface on mountain tops just beyond the terminator nevertheless
enabled us to obtain high signal-to-noise bright side spectra well
within the CCD detector's linear response regime.

The data reduction and extraction of the Earthshine spectra followed
the general technique and procedures of Woolf et al. (2002) and Arnold
et al.  (2002) who first reported the possible detection of the red
edge feature in an Earthshine spectrum.  Standard IRAF image
processing and spectral reduction routines were used with the
exception of sky subtraction for the dark limb exposures, as described
below.  First, standard bias frame subtraction and flat fielding
corrections were carried out for each exposure.  A sky contribution to
the dark limb spectra was determined by a linear extrapolation of the
observed sky counts at each wavelength to each position along the part
of the slit positioned on the lunar surface.  Visual examination of
the sky counts versus position along the slit beyond the lunar limb
established that the linear extrapolation was an adequate
representation of the spatial variation of sky brightness due to
scattering from the nearby (in the sky) bright lunar crescent.  Sky
subtraction for the bright side spectra was made in the conventional
way by subtracting an interpolation between the dark areas on either
side of the sunlit mountain peaks but was in any case so small as to
be insignificant.  The resulting dark limb spectra were then averaged
over about 30 arcsec of the slit's length across the lunar surface,
both to improve signal-to-noise and average out any fluctuations in
the wavelength dependence of the lunar albedo (i.e., changes in the
color of the lunar surface).  Unfortunately, no such averaging was
possible of the spectra of the lunar mountain tops, and as a result,
any difference between their wavelength dependent albedo and that of
the averaged areas observed on the dark limb will appear as systematic
noise, and perhaps a spurious feature, in our final Earthshine
spectra.  Finally, the spectra of the dark limb were divided by those
of the sunlit mountain peaks and averaged together to give a spectrum
of the Earthshine that illuminated the dark side of the Moon on each
of the two dates.  These are shown in Figure~\ref{fig:apo} along with
a representation of the Earth as it appeared from the
Moon\footnote{http://www.fourmilab.ch/earthview/vplanet.html} at the
times of the observations.  The nominal signal-to-noise ratio of these
spectra is in the range of 30 to 100 per resolution element with a
significant contribution from high spatial frequency CCD flat field
features, but the true uncertainty in the spectra is almost certainly
dominated by potential systematic errors discussed in the next
section.

\subsection{Interpretation of APO Earthshine Spectra}

The most important conclusion to be drawn from the spectra shown in
Figure~\ref{fig:apo} is that they appear to confirm the tentative
identification of the 700--750~nm feature, seen in the
Earthshine spectra of Woolf et al (2002) and Arnold et al.(2002), as
actually being due to the red edge in the spectral reflectance curves
of the leaves of deciduous terrestrial plants.  This conclusion is
based on the fact that a similar feature to the Woolf et al. (2002)
and Arnold et al. (2002) feature occurs in our ``vegetation-covered''
Earth spectrum but not in our non-vegetation-covered Earth spectrum.
In our vegetation covered spectrum from 8 Feb 2002 the Moon was
illuminated by light scattered from large land areas that are covered
by heavy vegetation---most notably the extensive forests of South
America. In our non-vegetation covered spectrum of 16 Feb 2002, the
Moon was primarily illuminated by light scattered off of parts of the
Earth's surface that contain no deciduous vegetation---primarily the
South Pacific Ocean.  Thus, at a crude and qualitative level at least,
the presence and absence of the red edge feature correlates with the
presence and absence of large quantities of plant life in the sunlight
illuminated portion of the Earth as seen from the Moon.

Spectral slope is routinely used to quantify detections of the red
edge via satellite observations of the Earth.  In the
vegetation-covered APO spectrum the continuum level of the Earthshine
rises by 17 +/- 2\% between 700 and 750 nm.  In contrast, the
non-vegetation-covered APO Earthshine spectrum has an increase of only
11 +/- 2\% over the same wavelength range\footnote{Turbull et al. (in
preparation) suggests that the gradual increase in slope of the
spectra throughout the entire wavelength range is due to the
difference in lunar albedo as a function of phase angle; the dark part
of the moon is illuminated by Earthshine at a phase angle roughly near
10 degrees whereas the bright crescent moon is illuminated by the sun
closer to 170 degrees.}  These values appear plausibly consistent with
the 6\% increase reported by Woolf et al. (2002) and of 4 to 10\%
reported by Arnold et al. (2002) in spectra of Earthshine produced by
light scattered from various other parts of the Earth's surface.

The change in the slope of the continuum in the vegetation-covered
case is even more striking; the slope is 14 +/- 3 times larger between
700 and 750 nm than between 650 and 700 nm!  This abrupt break
corresponds to the ramp-like appearance of the continuum at the red
edge location in the 8 Feb spectrum shown in Figure 4.  The
non-vegetation-covered spectrum also shows a mild reddening of the
continuum slope, by a factor of about 2.5 +/- 0.5, in a comparison of
the same two wavelength intervals.  This much smaller feature could be
due to a minor red edge contribution to Earthshine from plants near
the illuminated limb of the Earth (as seen from the Moon), but it is
not significantly bigger than changes in the continuum slope at other
wavelengths due to gentle bumps and dips in the spectrum's overall
shape. Thus we cannot claim secure detection of any feature at the red
edge location in the 16 Feb data.

Although a more quantitative comparison of the expected and observed
strength of the feature would be possible via calculations such as
those reported by Ford et al. (2001), we do not believe that the
quality and limited quantity of the currently available data justify
such a comparison.  However, it is clear that a more well-developed
data set of Earthshine spectra, taken at different moon phases and
times (seasons), as well as from different locations on the Earth's
surface, could unambiguously establish the association of the 750~nm feature with the red edge of vegetation, as well as document
its variability.  Such data would also be very useful for purposes of
testing and validating models of the Earth's photometric properties
such as that developed by Ford et al. (2001).

The spectra shown in Figure~\ref{fig:apo} display two additional
characteristics that merit comment: 1) a variation in the strength of
the water vapor lines between the 8 Feb and 16 Feb 2002 spectra and 2)
another strong increase in brightness around a wavelength of 850~nm.

Taking them in turn, the water vapor line variability is likely due to
differences in the average water vapor column density in the Earth's
atmosphere over the illuminated regions on the two different dates of
observation (see Figure~\ref{fig:apo}). We note that the water vapor
features do not show any systematic variation with airmass in subsets
of the spectra obtained on either date.  We caution, however, that
there is a small chance that
the water variation may nevertheless reflect imperfect cancellation or
removal of the atmospheric water vapor in front of the telescope
during the observations due, perhaps, to the high and rapidly changing
airmass through which the observations were necessarily obtained.

We have found no plausible explanation for the 850~nm feature; it
could be real and deserving of further investigation.  It might also be
artificial in the sense of arising not on Earth but from the Moon; in
other words it might be due to a variation in the color of the lunar
surface between the regions where the dark limb and the sunlit surface
spectra were obtained, as described above in \S\ref{sec-observations}.
Any such difference would not be removed by the division of the dark
by the bright side spectra.  We note, however, that both Earth-based
(McCord et al.  1972) and
spacecraft\footnote{http://vims.artov.rm.cnr.it/data/res-moo.html}
measurements of the wavelength dependence of the albedo of lunar
highland regions do not show any strong features at either the
position of the red edge or near 850~nm.  Whether the 850~nm
feature is created at the Earth or on the Moon, it warns of the
difficulties that may arise when searching for red edge-like spectral
features on other planets. We note that the Earthshine spectrum by
Woolf et al. (2002) shows a 7\% decrease (approximately three times
smaller change than we see at 850~nm) at wavelengths
$\sim$~780~nm and longer. Turnbull et al. (in preparation) also
see a depression in their reflectance spectra at similar wavelengths.

These systematic uncertainties are not only relevant to the particular
data presented here but also illustrate some of the intrinsic
challenges of Earthshine observations using general purpose telescopes
and instruments.  The sunlit and Earthshine illuminated portions of
the lunar disk differ in surface brightness by at least on the order of
$10^4$ and are thus very difficult to observe with comparable
signal-to-noise using the same equipment and instrumental
configuration.  Moreover, rapidly changing airmass and a very limited
period of time available for the whole sequence of exposures are
unavoidable consequences of observing a crescent moon from
observatories at mid-latitudes.  These time constraints make it
impractical to reconfigure the instrumentation repeatedly to
facilitate observations of the bright and dim portions of the lunar
surface, {\it e.g.}, by insertion and removal of a pupil plane
diaphragm or a neutral density filter in a spectrograph that was not
designed with such operations in mind.  And, of course, the high
airmasses associated with crescent moon configurations are directly
problematic; they change significantly over a few minutes and thus
even during and between fairly short exposures.  Earthshine
spectroscopy using special purpose instruments or instrument
configurations either from polar regions of the Earth (Kilston,
private communication, 2002) or orbiting telescopes (Davis et al. 2002;
Woolf,
private communication, 2001) could circumvent these difficulties.  

\section{Extrasolar Plants}
\label{sec-extrasolarplants}

The red edge will only be a useful tool for astrobiology if extrasolar
vegetation also exhibit a similar extraordinary spectral signature.
We have no firm basis for believing in the existence of extrasolar
vegetation. It is nevertheless essential to keep in mind that the
situation is qualitatively the same for every technique that can be
used for the search for life in the Universe.  Absent a fundamental,
first principles understanding of biology that could predict what is
possible and likely, we are forced to assume that extraterrestrial
life resembles life on Earth {\it to some degree}. Assuming that
extraterrestrial life has {\it nothing at all in common} with
terrestrial organisms makes searching for it almost impossible.  The
opposite extreme of assuming that terrestrial and extraterrestrial
life are {\it exactly identical} is simplistic and would lead to such
narrowly focused search techniques that would risk missing most actual
indications of extraterrestrial biology.

In this section we offer some of the relevant facts and arguments to
inform the required educated guess.  This is in contrast to
Wolstencroft and Raven (2002) who use details of the photosynthesis
mechanism to argue that the evolution of O$_2$-producing
photosynthesis is likely on extrasolar Earth-like planets.
A few interesting points concerning terrestrial vegetation are useful
for speculating on the possible existence of extrasolar plants or
``light harvesting organisms'' and potential magnitude of spectral
features:


$\bullet$ Plants absorb very strongly throughout the UV (at
wavelengths longer than 380~nm) and the visible wavelength
regions of the spectrum where photon energies are sufficient to drive
photosynthesis (involving molecular electronic transitions);

$\bullet$ At sea level (i.e., after atmospheric extinction) the solar
energy distribution peaks at 1000~nm and approximately 50\% of the
energy is redward of 700~nm (Figure~\ref{fig:atm}). Plants reflect
and transmit almost 100\% of light in the wavelength region where the
direct sunlight incident on plants has the bulk of its energy (Gates et al. 1965);

$\bullet$ Considering the above two points, Earth's primary surface
``light harvesting organism'', vegetation,  may have evolved to
balance the competing requirements of absorption of sunlight at
wavelengths appropriate for photosynthesis reactions with efficient
reflectance at other wavelengths to avoid overheating (Gates et
al. 1965). The prime selective factor in evolution, however, is
not known, and vegetation's high reflectance and transmittance may
instead have been selected for other traits, namely the large
intercellular gas spaces aiding gas exchange (Konrad et al. 2000) and
increasing the absorption of photosynthetically active radiation
(DeLucia et al. 1996). Regardless of the evolutionary origin, the high
reflection may be key to vegetation's surival on land.

$\bullet$ In addition to chlorophyll (Chl), many other light-absorbing
pigments exist.  Accessory pigments at different wavelengths
(Figure~\ref{fig:pigments}) absorb photons and transfer energy to Chl
a. These accessory pigments allow photosynthetic organisms to use a
wide range of wavelengths of light.  Specifically: Chl a is found in
all photosynthetic organisms except some photosynthetic bacteria; Chl
b is found in higher plants and green algae; accessory pigment $\beta$
carotene is found in all photosynthetic organisms except
photosynthetic bacteria; phycoerythrin and phycocynanin (phycobilins)
are found in red algae and cyanobacteria respectively. In addition to
the variety of accessory pigments, the chlorophyll pigment itself
comes in different forms with absorption maxima at different
wavelengths. For example, photosynthetic bacteria (both aerobic and
anaerobic) have bacteriochlorophyll (Bchl) pigments
(Figure~\ref{fig:pigments}). The wavelength variation in the pigments
makes some organisms better adapted to their ecological niches. For
example, cyanobacteria (Chl a and phycocyanin) in lakes and ponds
often form a dense surface layer, absorbing a large amount of blue and
red light.  Purple photosynthetic bacteria (Bchl a or b) and green
photosynthetic bacteria (major pigment is Bchl c, d, or e) grow best
in anaerobic conditions in deep water.  At depth, the previously
cyanobacteria-absorbed blue and red light is not available.  The
bacteriochlorophyll pigments allow the purple and green bacteria to
take advantage of their ecological niche in deep water by absorption
of longer wavelength light. In addition, Bchl a and b have absorption
maxima at shorter wavelengths than Chl a, taking advantage of the deep
water where shorter wavelength light can penetrate water farther.

$\bullet$ While vegetation and other organisms use chlorophyll
pigments to convert light to energy, there exists at least one 
alternate photosynthetic system using an independent pigment,
rhodopsin. Bacteriorhodopsin (a type of rhodopsin) occurs in 
halobacteria (Oesterhelt \& Stoeckenius 1971) found
in highly salty environments, for example in the Dead
Sea. Proteorhodopsin is found in marine bacterioplankton that are
widespread in the surface ocean (Beja et al. 2001). The rhodopsin
photosynthetic system is chemically fundamentally different from the
chlorophyll photosynthetic system.

If we accept that extraterrestrial light-harvesting organisms should
be ubiquitous, would they likely have a strong spectroscopic
signature?  Light harvesting organisms may have similar properties to
vegetation in order to absorb the correct frequency energy for
molecular transitions but not absorb all available energy.
Light-harvesting pigments in vegetation cover the full range of the
visible-light spectrum and many of these pigments have sharp spectral
features at the red edge of the pigment's absorbing range
(Figure~\ref{fig:pigments}).  Thus, extrasolar light harvesting
organisms may have sharp spectral features similar to terrestrial
vegetation's red-edge spectral feature but at different wavelengths.

In some cases light harvesting organisms might have no strong
spectroscopic signature. For example, a hypothetical light-harvesting
organism that is absorptive at short visible wavelengths but purely
transmissive at red and near-IR wavelengths would have no sharp
spectroscopic features detectable by reflectance spectroscopy.  A
second example comes from terrestrial photosynthetic organisms. In
addition to vegetation there is a large diversity of photosynthetic
organisms (e.g., multi-cellular and unicellular algae and prokaryotic
organisms such as cyanobacteria and green and purple bacteria). Over
half the photosynthesis on Earth is carried out by microorganisms
(Field et al. 1998). Many of these organisms live in water and some
carry out anoxygenic photosynthesis---any spectral biosignature would be
weakened in the spatially integrated global spectrum due to the high
opacity of ocean water.

We favor the opinion that light harvesting organisms with pigments and
spectral features at various wavelengths should be common over the
idea that the red edge signature from terrestrial plants is widespread
and universal (although the pigment ``edge'' may or may not be a sharp
feature; see Figure 5).  Given the abundance and ready availability of
low-entropy energy in the form of radiation from the primary star,
assuming a sufficiently transparent atmosphere, ``light-harvesting" is
likely to be a common feature of life, simply because it is such an
effective biological strategy.  In other words, if such a beneficial
mechanism develops in any organism, it seems likely to be subject to
strong positive evolutionary selection.  Wolstencroft \& Raven (2002)
also conclude that light harvesting organisms with pigments and
spectral features at different wavelengths should be common, by the
different assumption that Earth-based photosynthesis should be common.

Evolution has an element of chance, and whatever light-harvesting
mechanism develops first on a planet might be evolutionarily favored
over other mechanisms that are theoretically more efficient.  Thus,
Earth-based photosynthesis is not necessarily the best or most
efficient light-harvesting mechanism even for the conditions on Earth.
Further, there is no compelling {\it a priori} reason to believe that
organisms on other planets would independently develop a
light-harvesting mechanism identical to the one found in terrestrial
vegetation.  Furthermore, many mechanisms evolved initially for other
functions.  For example, it has been suggested that chlorophyll 
arose from UV-screening, cell-surface proteins (Mulkidjanian \& Junge
1997). The proteins that originally developed as a protective measure
later evolved into using the absorbed radiation for energy.  So, for
example, organisms living on planets around stars cooler than the sun
may have pigments at slightly different wavelengths (see Wolstencroft
\& Raven 2002 for a discussion of O$_2$-producing photosynthesis
around cool stars).

Our chain of reasoning does not lead to any definite conclusion;
however it does give us some basis for an {\it opinion}: a
red-edge-like spectroscopic signature, as a biosignature
for astrobiology, is sufficiently promising
to warrant the search for similar features
in the spectra of extrasolar terrestrial planets using TPF-type
and successor missions.  If such a feature were detected with time
variability and other systematic behaviors, we believe that it would
be an extremely interesting clue that suggests the possible existence
of light-harvesting organisms on the planet.  On the other hand, we do
not believe that a failure to detect a red-edge-like signature in the
spectrum of an extrasolar terrestrial planet would provide any
meaningful null result with respect to life on the planet.

\section{False Positive Mineral Reflectance Edges}
\label{sec-minerals}
Semiconductor crystals also have spectral reflectance ``edges'' at or
near visible wavelengths (Figure~\ref{fig:minerals}).  The spectral
reflectance edges are due to the valence electrons lacking available
states for a certain range of energies.  The semiconductor's band gap
is the energy difference between the valence shell band and the
conduction band.  Photons with enough energy are absorbed, since they
can excite electrons from the valence band into the conduction band.
Photons with energies less than the band gap, however, are absorbed and
remitted with the same energy---i.e. reflected---by electrons within
the valence band.  At zero temperature, there is a sharp step function
reflectance edge at the wavelength of the energy gap.  At higher
temperatures the reflectance edge is smooth and sloped, since there is
a chance that an incident photon will excite an electron already above
the ground state and the combined energy will be sufficient for the
electron to jump the band gap.  For many semiconductors, the band gap
energy corresponds to the energy of photons in the visible to near-infrared
portions of the spectrum.  Therefore, mineral semiconductors will also
have spectral reflectance edges at visible to near-infrared wavelengths.
For example, cinnabar (HgS) is a mineral with a steep reflectance edge
(Figure~\ref{fig:minerals}) at red wavelengths (600~nm).  It is
important to remember that the band gap phenomena is dependent on the
crystalline structure of the semiconductor which creates a periodic
potential for the impinging photon.  Thus, if the physical size of the
crystal does not significantly exceed the wavelength of the incident
photons, then the spectra can be significantly altered by the shape
and size of the crystals.  In the solar system, most planetary
surfaces are covered with a regolith that complicates the
interpretation of mineral reflectance spectra.

A planet with a lot of exposed rocks (aggregates of mineral) could 
produce a strong mineral-edge signature, even if the planet had an 
atmosphere that was optically thin at visible wavelengths.   Like the 
vegetation edge, a mineral reflectance edge could vary as continents 
rotate in and out of view.  A mineral edge detection would be 
interesting in and of itself.   For example, a substantial fraction of 
Jupiter's moon Io is likely covered by solidified elemental sulfur 
allotropes (Moses \& Nash 1991) that originated from volcanic eruptions. 
Crystalline sulfur is an an intermediate energy band gap semiconductor 
and has a sharp reflectance edge at 0.45 microns (Figure~\ref{fig:minerals}). 
Disk-averaged reflectance spectra of Io are remarkable in that blueward 
of the mineral edge Io's albedo is near zero and redward of the mineral 
edge the albedo is 0.8 (e.g., Spencer \& Schneider 1996).

If a reflectance edge were detected in the spectrum of an extrasolar
terrestrial planet, then it would be important to consider all
possible sources, including both mineral and biological.  In
particular, a careful study of mineral signatures and atmospheric
composition will be necessary before attributing a spectral edge to a
light-harvesting pigment.  Although only a dozen or so rock-forming
minerals are common on Earth, over four thousand minerals are known.
While their spectra can be measured in the lab, and compared with the
extrasolar planet reflectance spectrum, it can be tricky to identify a
particular mineral from a reflectance spectrum (e.g., Moses \& Nash
1991.)  Additionally, the spectra surfaces covered with a regolith of
small particulate minerals can deviate significantly from laboratory
spectra based on large crystals.  Nevertheless, it may be possible to
eliminate the possibility of some minerals based on additional
measurements.  For example, most minerals have near-IR and mid-IR
absorption features.  Therefore, a mid-IR spectrum (e.g., from the
mid-infrared {TPF/Darwin) at wavelengths where radiation penetrates to
the planet's surface could help identify a mineral that covered a
large fraction of a planet's surface.  Measurements of atmospheric
composition could also be valuable for ruling out certain minerals.
For example, if a detected spectrum showed a significant amount of
atmospheric oxygen, then non-oxidized minerals are unlikely to be
abundant on the planetary surface. The wavelength of the Earth's
vegetation red edge does not correspond to that of any known mineral
(Sagan et al. 1993).

\section{Summary and Conclusions}

When extrasolar Earth-like planets are discovered, observations at
wavelengths that penetrate to the planet's surface will be useful
for detecting surface features, including biosignatures, especially
for planets with much lower cloud cover than Earth's 50\%.  The
vegetation red edge spectroscopic feature is a factor of 5 or more
change in reflection at $\sim$ 700~nm in deciduous plant
leaves. This red edge feature is often used in remote sensing studies
of Earth's vegetation. Earthshine observations have detected a feature
identified with the red edge in the spatially unresolved spectrum of
Earth where it appears at the few percent level.  The Earthshine
observations reported here display the expected dependence of the
feature on the portion of the Earth both illuminated by sunlight
and visible from the Moon and thus support this interpretation.

Earth's hemispherically integrated vegetation red-edge signature,
however, is weak (a few to ten percent vs. the $\gtrsim$ 50\% in leaf
reflectance) due to dilution by other atmospheric (e.g., clouds) and
surface (e.g., oceans and non-continuous forest coverage) features.
The recent Earthshine measurements have confirmed the expected
amplitude of the red-edge signature, and show that its detection is
non-trivial in Earth's spatially unresolved spectrum with current and
near-future planned technology.  Earth-like planets with different
rotation rates, obliquities, higher land-ocean fraction, different
continental arrangement and lower cloud-cover might well display a
more easily detectable red-edge-type signal.

A time series of spectra or broad-band photometry will help to
identify weak surface biosignatures such as the red-edge-type feature in
a spatially unresolved spectrum of an extrasolar planet.  The new
Earthshine observations reported here indicate just such temporal
variability of Earth's red edge surface biosignature.  The increased
temporal variability at a carefully chosen color could make detection
of such features easier.  In particular, any changes associated with a
rotational period would be highly relevant, but
c.f. \S\ref{sec-minerals} for a discussion of surface mineral false
positives.

Because the existence of extraterrestrial light harvesting organisms
is plausible, any prediction of the characteristics of light
harvesting organisms on extrasolar planets is highly speculative and
uncertain. In particular, the wavelength of any surface biosignatures
would not be known {\it a priori}.  Therefore, flexible data
acquisition will maximize scientific return from future missions. For
example, spectra can be integrated into photometry in many possible
different bands after the data is acquired.  Similarly, a relatively
high cadence of observations can later be searched for rotational
periods associated with surface features.

The physical and evolutionary characteristics of the red edge and the
biological significance of photosynthesis make it plausible that
extraterrestrial organisms might use pigments to harvest light, and
that these organisms might develop a red-edge-like spectral feature to
protect their pigments.  If a red-edge-like feature were detected in
combination with the spectral signatures of biosignature gases, then it
could be strongly suggestive of the presence of life on the planet.
Moreover, the detection of {\it any} unusual spectral feature that is
inconsistent with known atomic, molecular, or mineralogical signatures
would be extremely interesting, and the fact that the Earth's spectrum
displays at least one such detectable feature of biological origin is
encouraging.  Combinations of unusual spectral features together with
strong disequilibrium chemistry would be even more intriguing and
would certainly motivate additional studies to better understand the
prospects for such a planet to harbor life.

\begin{acknowledgments}
We thank the members of the Princeton Terrestrial Planet Finder group
plus Steve Kilston, Wes Traub and Nick Woolf for valuable discussions
of the red edge and Earthshine. We also thank Mike Wevrick for useful
contributions, Harold Morowitz, Andrew Steele, Jan Toporski, Maggie
Turnbull, and Nancy Kiang for useful discussions about chlorophyll and
leaf structure, Olga Degtyareva for translating Arcichovsky (1912),
Tom Murphy for providing specialized lunar pointing and tracking
telescope control software, Merri Wolf for careful library research,
and Gabriela Mallen-Ornelas for contributing APO observing time and
for a careful reading of the manuscript.  Russet McMillan and Camron
Hastings of APO supplied expert assistance with the somewhat
unconventional demands of Earthshine observations using a large
telescope and a faint object spectrograph.  We thank the referees for
valuable comments.  S. S. is supported by the Carnegie Institution of
Washington. This material is also based upon work supported by the
National Aeronautics and Space Administration through the NASA
Astrobiology Institute under Cooperative Agreement NCC 2-1056.  Work
at Princeton University on this topic is supported in part by NASA
grant NAG5-13148.
\end{acknowledgments}

\begin{figure}
\plotone{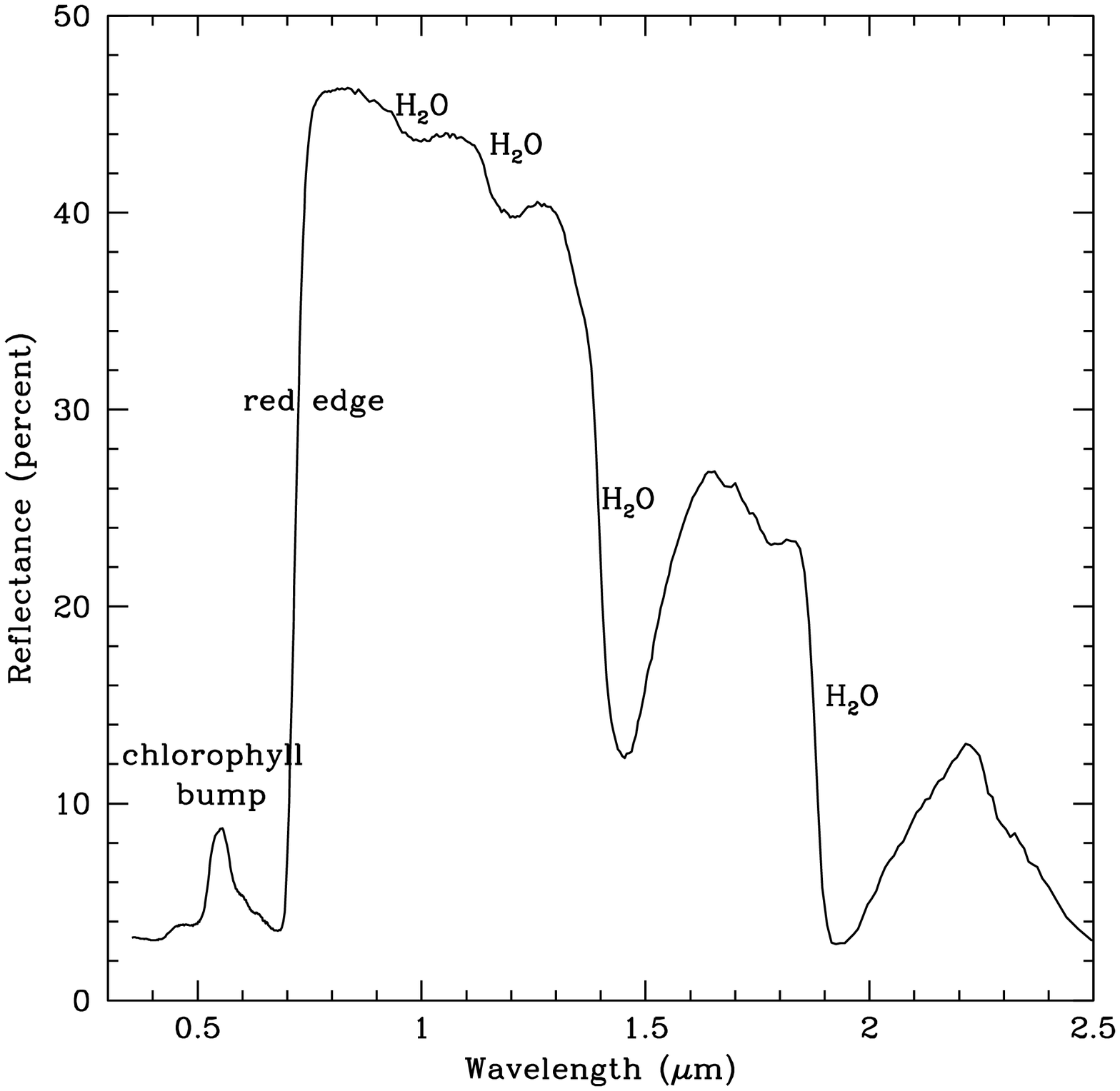}
\caption{Reflection spectrum of a
  deciduous leaf (data from Clark et al. 1993). The small bump near
  500~nm is a result of chlorophyll absorption (at 450~nm and
  680~nm) and gives plants their green color. The much larger
  sharp rise (between 700 and 800~nm) is known as the red edge and
  is due to the contrast between the strong absorption of chlorophyll
  and the otherwise reflective leaf.}
\label{fig:plantspectrum}
\end{figure}

\begin{figure}
\plotone{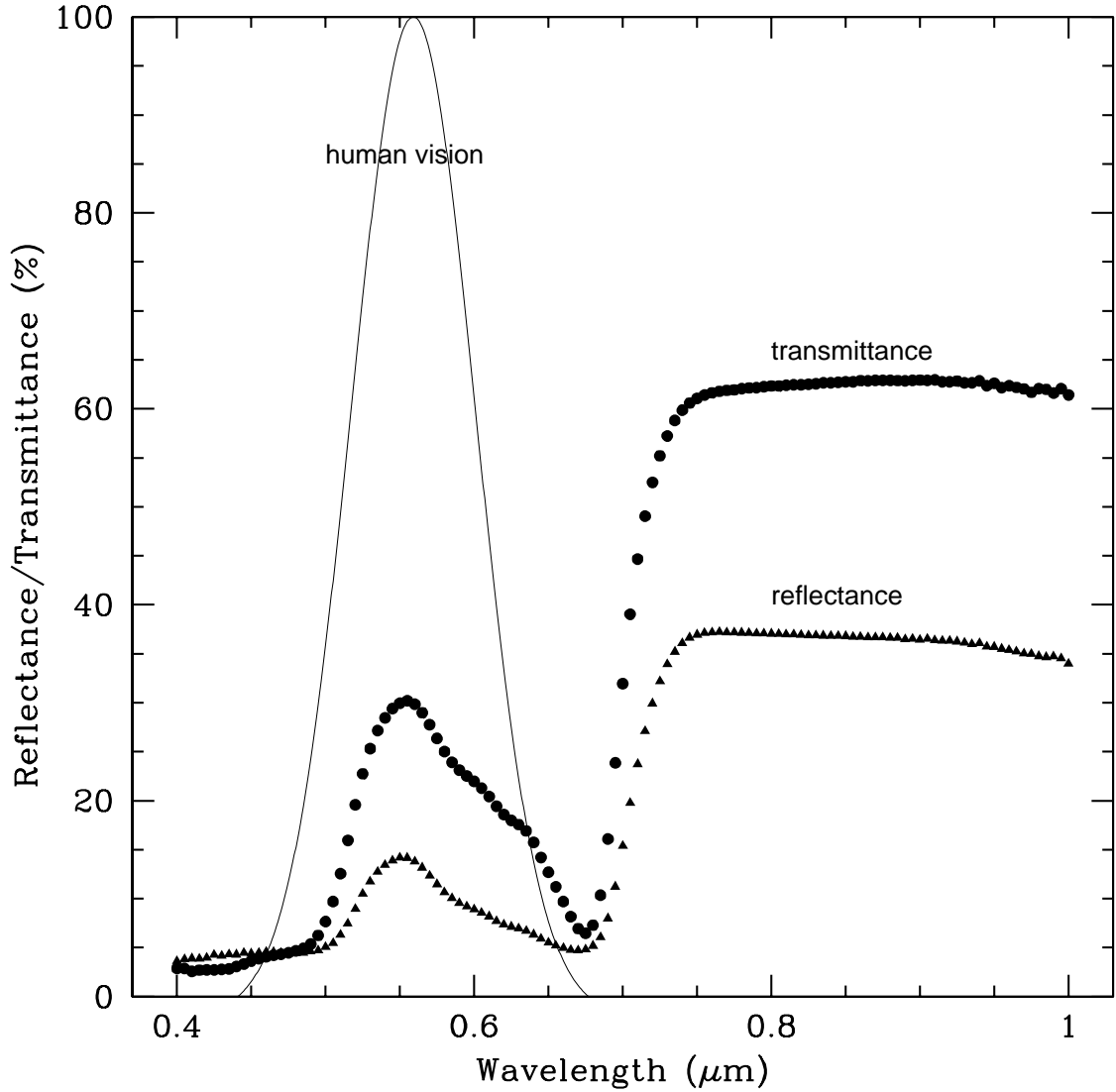}
\caption{ Reflectance and transmittance of a young aspen leaf (data
from Middleton \& Sullivan 2000). The reflection and transmittance
curves are very similar, supporting the point that radiation keeps
scattering until it exits the top surface or bottom surface of the
leaf, except at pigment-absorbing wavelengths. Also shown is the
normalized human vision response curve which shows the nonsensitivity
of humans to the vegetation red edge reflectance.}
\label{fig:plantcurve}
\end{figure}

\begin{figure}
\plotone{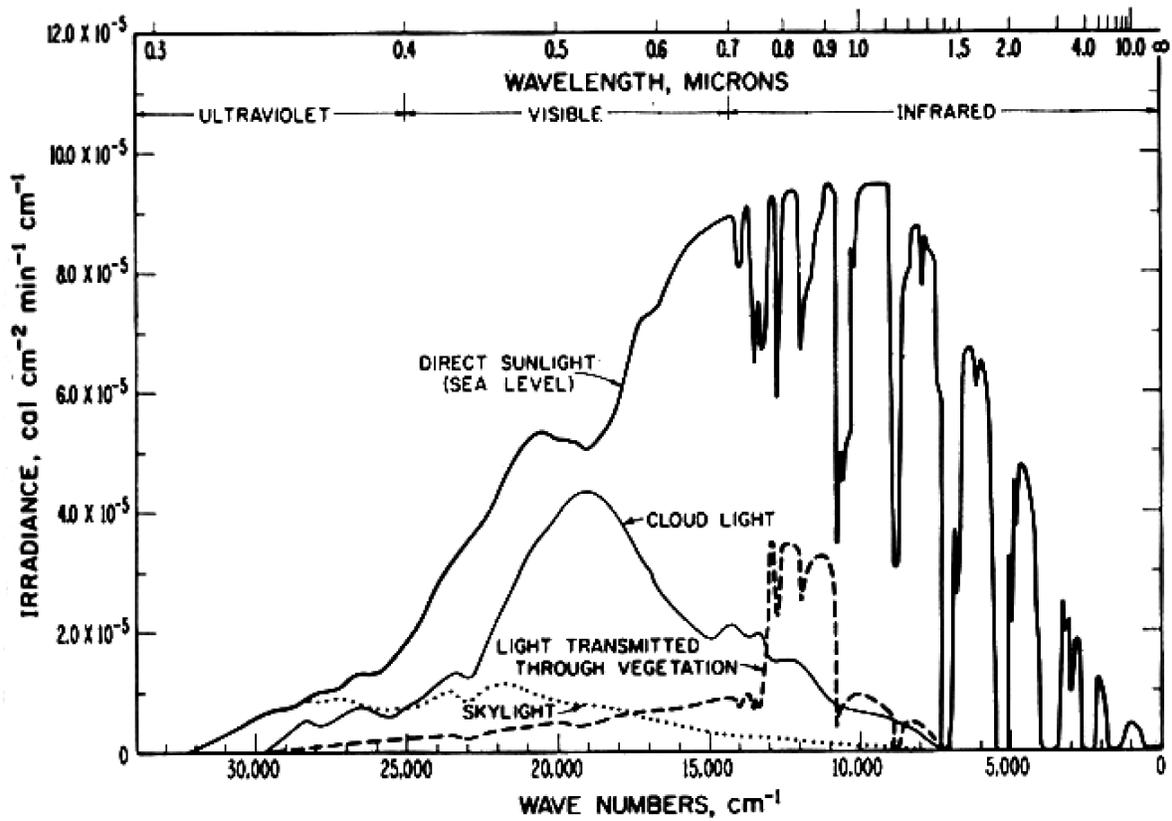}
\caption{Spectrum of solar flux that reaches the ground after
absorption through Earth's atmosphere. Also shown is a representative
curve for light transmitted through vegetation (light is reflected at
the same wavelengths; see Figure~\ref{fig:plantcurve}).  Reproduced
from Gates et al. (1965). Vegetation reflects or
transmits almost all incident radiation at wavelengths where
atmosphere-filtered sunlight has the bulk of its energy.}
\label{fig:atm}
\end{figure}

\begin{figure}
\plotfiddle{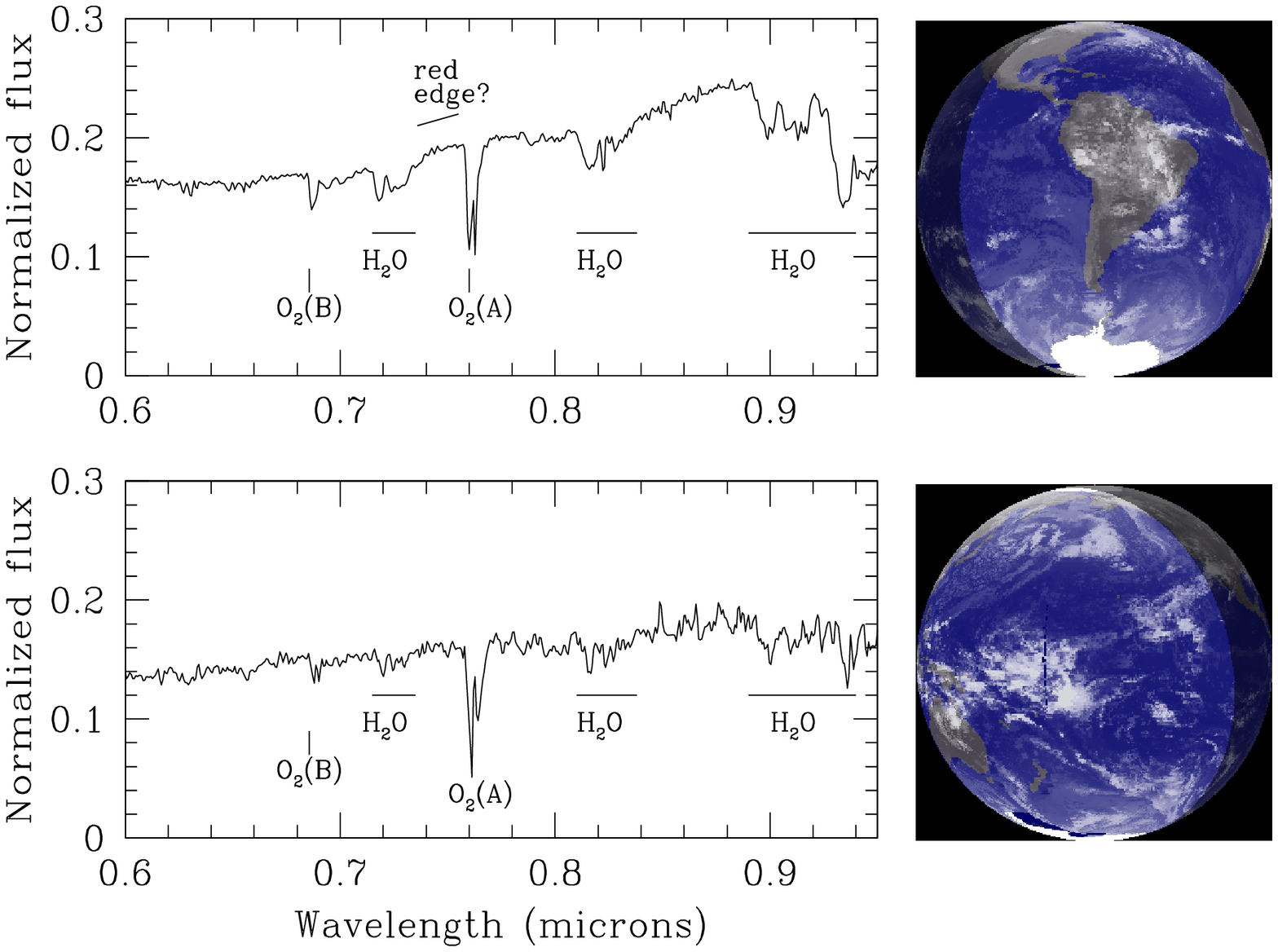}{5.2in}{0}{75}{75}{-230}{-150}
\plotfiddle{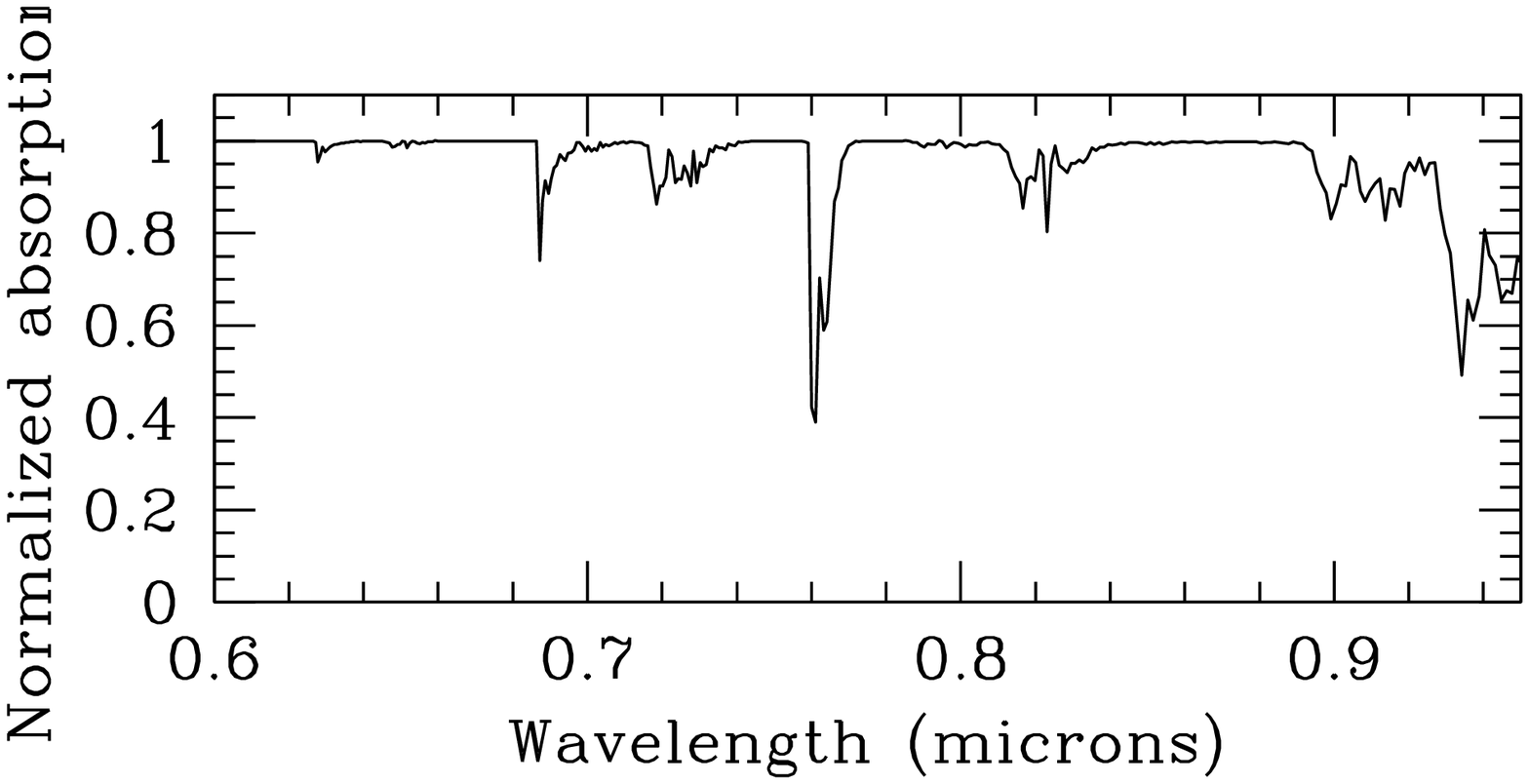}{.5in}{0}{47.5}{56}{-180}{-226}
\caption{Earthshine observations from Apache Point Observatory. Top
panel: Earthshine observations on 8 February 2002.  The viewing
geometry (including cloud coverage at the time of observations) of
Earth from the moon is shown in the right image
(http://www.fourmilab.ch/earthview/vplanet.html).  Middle panel: same
as upper panel for 16 February 2002. The viewing geometry of Earth
includes much more vegetation in the top panel than in the middle
panel. Bottom panel: an absorption spectrum through Earth's atmosphere
from Kitt Peak National Observatory
(ftp://ftp.noao.edu/catalogs/atmospheric\_transmission/) smoothed to
approximately the same resolution as the Apache Point Observatory
Earthshine data. Note the different y axis on the absorption spectrum;
the spectral features are much deeper than in the Earthshine spectra
and there is no red edge feature.}
\label{fig:apo}
\end{figure}

\begin{figure}

\plotone{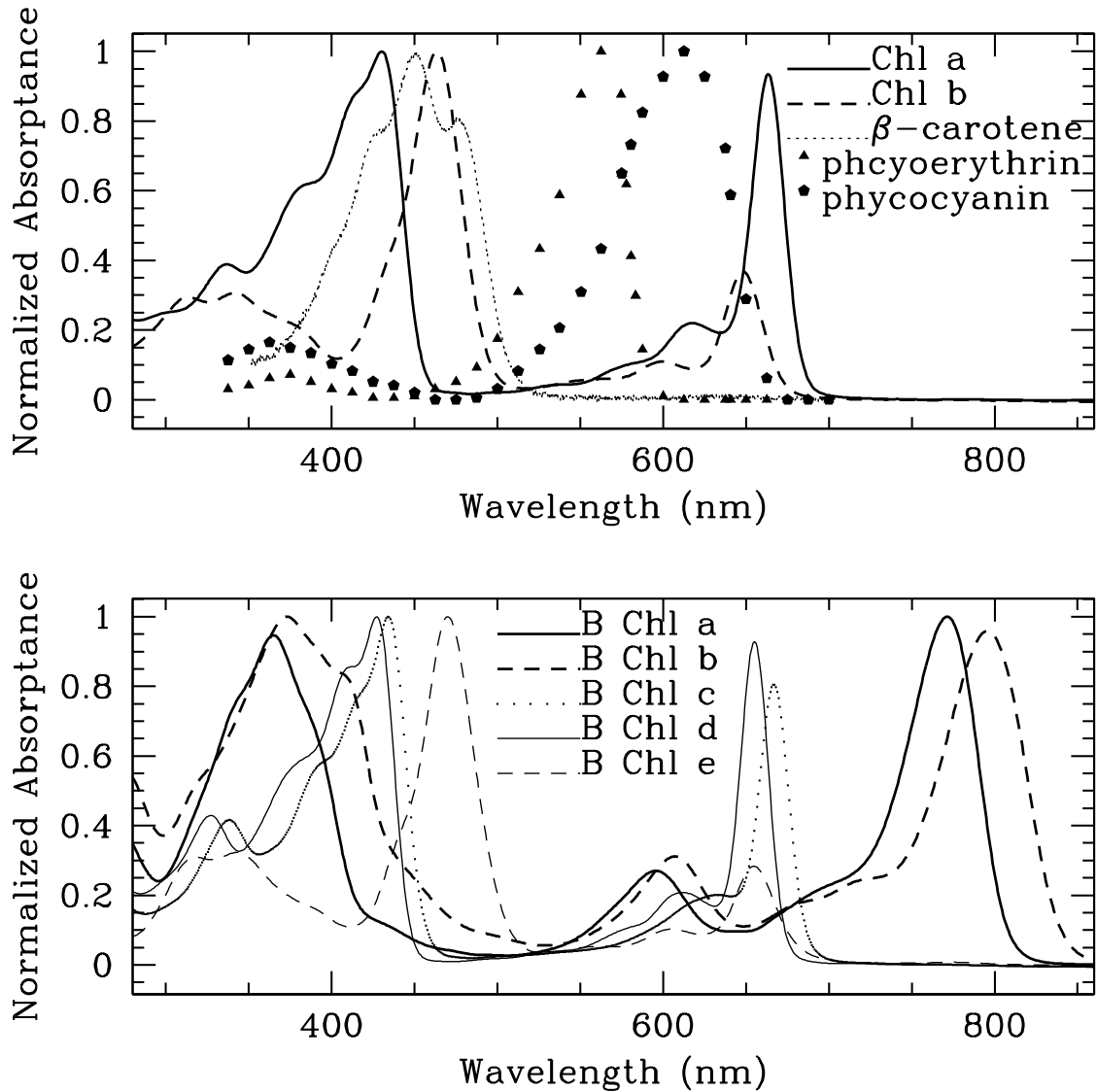}
\caption{Wavelength dependence of some of the pigments involved in
photosynthesis. Top: Chlorophyll a and b (dissolved in a solution of
methanol, acetonitrile, ethylacetate, and water; Frigaard et
al. 1996); accessory pigment $\beta$ carotene
(dissolved in hexane; Du et al. 1998); accessory pigments
phycoerythrin and phycocynanin (dissolved in water; Purves et al. 1995).
Bottom: Wavelength dependence of bacteriochlorophyll pigments. The
bacteriochlorophylls are found in photosynthetic bacteria. The data
plotted here are pigments dissolved in a solution of methanol,
acetonitrile, ethylacetate, and water. Data courtesy of Niels-Ulrik
Frigaard (Frigaard et al. 1996).  See Section
\ref{sec-extrasolarplants} for discussion.}
\label{fig:pigments}
\end{figure}

\begin{figure}
\plotone{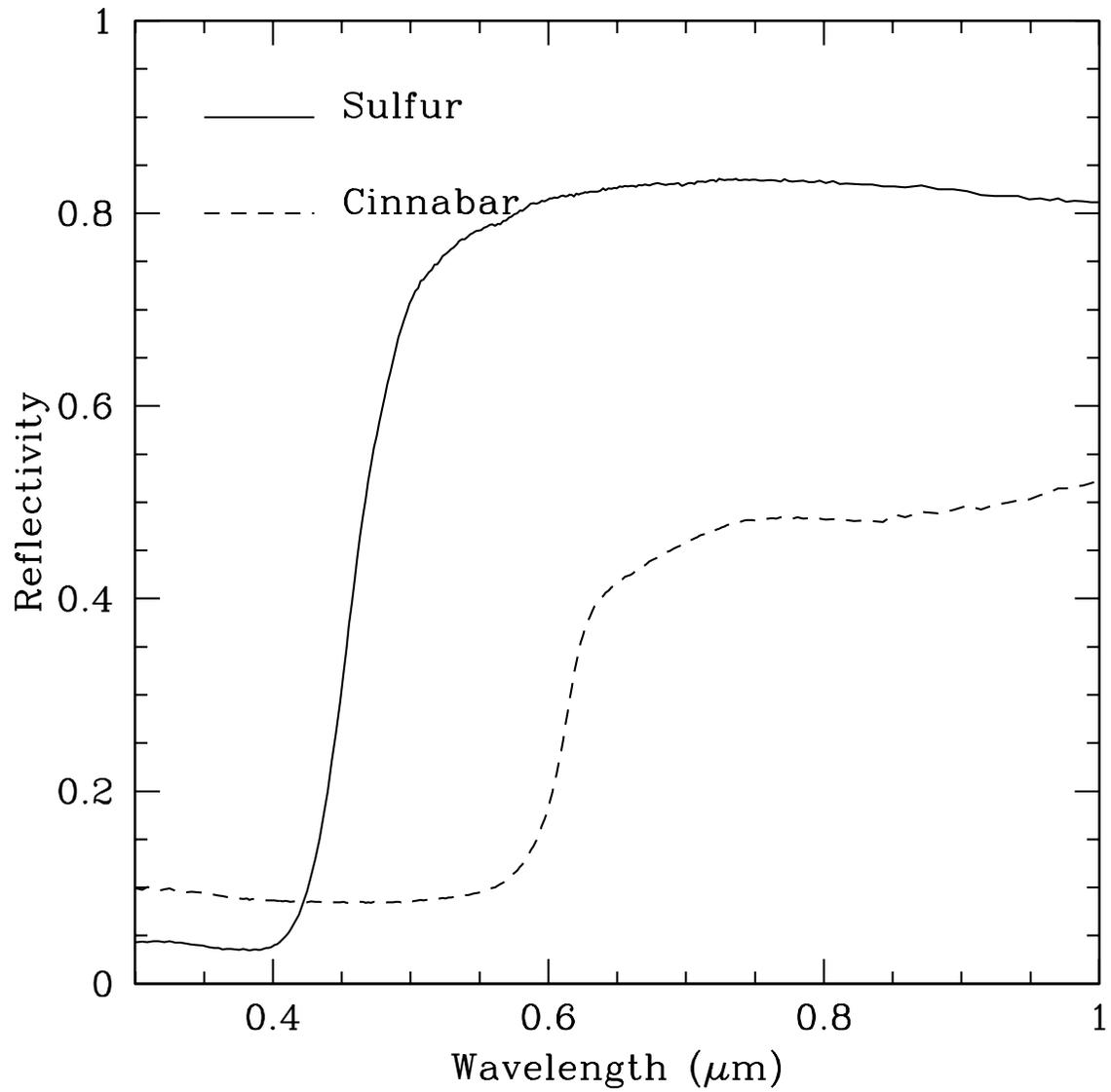}
\caption{Reflectance spectra of the minerals sulfur and cinnabar (from
Clark et al. 1993).  Note the sharp edge features.}
\label{fig:minerals}
\end{figure}

\end{document}